\documentclass[11pt]{IEEEtran}
\pdfoutput=1
\usepackage[utf8]{inputenc}
\usepackage[margin=1in]{geometry}
\usepackage{amsmath,amssymb,amsthm}
\usepackage{graphicx}
\usepackage{hyperref}
\usepackage{booktabs}
\usepackage{enumitem}
\usepackage{acronym}
\usepackage{orcidlink}    % \orcidlink{0000-0000-0000-0000} -> icono + enlace
\usepackage[utf8]{inputenc}
\usepackage[T1]{fontenc}

\acrodef{MAS}[MAS]{Multi-Agent System}
\acrodef{AI}[AI]{Artificial Intelligence}
\acrodef{BFT}[BFT]{Byzantine Fault Tolerance}
\acrodef{PBFT}[PBFT]{Practical Byzantine Fault Tolerance}
\acrodef{LLM}[LLM]{Language Large Model}
\acrodef{SECP}[SECP]{Self Evolving Coordination Protocol}

\title{Towards Selection as Power: Bounding Decision Authority in Autonomous Agents}
\date{}
\author{%
  \begin{tabular}{@{}c@{\hspace{1cm}}c@{}} % ajustar \hspace{...} para separación horizontal
    \begin{minipage}[t]{0.42\textwidth}\centering
      Jose Manuel de la Chica Rodriguez \small\orcidlink{0009-0009-9649-5805}\\
      {\small{Head of AI Lab}}\\
      {\small\textit{AI Lab, Grupo Santander}}\\
      {\small\textit{Madrid, Spain}}
    \end{minipage}
    &
    \begin{minipage}[t]{0.42\textwidth}\centering
      Juan Manuel Vera Díaz \small\orcidlink{0000-0002-6152-5789}\thanks{Corresponding author:\\ \texttt{juanma.vera@gruposantander.com}}\\
      {\small{Senior AI Researcher}}\\
      {\small\textit{AI Lab, Grupo Santander}}\\
      {\small\textit{Madrid, Spain}}
    \end{minipage}
  \end{tabular}%
}

\begin{document}

\maketitle

\begin{abstract}

Autonomous agentic systems are increasingly deployed in regulated, high-stakes domains where decisions may be irreversible and institutionally constrained. Existing safety approaches emphasize alignment, interpretability, or action-level filtering. We argue that these mechanisms are necessary but insufficient because they do not directly govern \emph{selection power}: the authority to determine which options are generated, surfaced, and framed for decision.

We propose a governance architecture that separates cognition, selection, and action into distinct domains and models autonomy as a vector of sovereignty. Cognitive autonomy remains unconstrained, while selection and action autonomy are bounded through mechanically enforced primitives operating outside the agent’s optimization space. The architecture integrates external candidate generation (CEFL), a governed reducer, commit--reveal entropy isolation, rationale validation, and fail-loud circuit breakers.

We evaluate the system across multiple regulated financial scenarios under adversarial stress targeting variance manipulation, threshold gaming, framing skew, ordering effects, and entropy probing. Metrics quantify selection concentration, narrative diversity, governance activation cost, and failure visibility.

Results show that mechanical selection governance is implementable, auditable, and prevents deterministic outcome capture while preserving reasoning capacity. Although probabilistic concentration remains, the architecture measurably bounds selection authority relative to conventional scalar pipelines. 

This work reframes governance as bounded causal power rather than internal intent alignment, offering a foundation for deploying autonomous agents where silent failure is unacceptable.

\noindent\textbf{Keywords:} 
\textbf{\textit{Constrained Optimization, Mechanism Design for AI Systems, Adversarial Robustness, Multi-Objective Decision-Making, AI Governance and Safety.}}

\end{abstract}

\section{Introduction}

\subsection{Why Selection Must Be Governed in Regulated Domains}

Autonomous agentic systems are increasingly deployed in regulated, high-stakes domains such as finance, healthcare, and critical infrastructure. In these environments, decisions frequently have irreversible consequences and are subject to explicit legal, ethical, and operational constraints. As a result, system failures are not merely technical malfunctions but governance failures with direct institutional and societal impact. The emerging literature on AI governance highlights the need for robust governance frameworks that ensure responsible AI deployment across the lifecycle of systems in such contexts \cite{batool2025aiGovernanceReview, monteiro2025wheel}.

Much of the existing literature implicitly assumes that unsafe or undesirable behavior stems primarily from insufficient reasoning capacity, misaligned objectives, or lack of internal oversight. Accordingly, substantial effort has been invested in improving cognition, alignment techniques, and interpretability. While these efforts are necessary, they are insufficient to explain or prevent a class of failures observed in practice, particularly when risk arises \emph{before} action execution.

In practice, many critical failures do not originate in what an agent computes, but in what it is allowed to decide among. Systems may comply with formal constraints, produce plausible explanations, and satisfy procedural requirements, yet still systematically steer outcomes away from governance intent. This occurs not through explicit rule violation, but through control over upstream processes that determine which options exist and how they are framed for decision-makers. This paper posits that in many regulated domains involving irreversible actions, \emph{selection power} constitutes a dominant source of institutional risk.

\subsection{Observed Failure Modes}

Across incidents and governance analyses, several recurrent failure modes emerge that are not explained by prediction error alone:

\begin{itemize}
    \item \textbf{Selection by exclusion (selection-by-negation).} Systems avoid generating or presenting options that are valid but reduce the agent’s expected utility or preferred outcome.
    \item \textbf{Framing manipulation.} The ordering or emphasis of rationales shifts human preferences without altering the underlying option set.
    \item \textbf{Variance inflation and reducer exploitation.} Score distributions are manipulated to destabilize aggregation mechanisms.
    \item \textbf{Entropy prediction and timing attacks.} Agents exploit randomness predictability to bias evaluations.
\end{itemize}

These behaviors operate upstream of action execution, escaping governance mechanisms that focus on final outputs only and not on the generation of choices.

\subsection{Operational Definitions}

To avoid ambiguity, we adopt the following definitions:

\paragraph{Selection Power} The capacity of an agent to influence outcomes by (i) shaping the candidate option space, (ii) controlling the subset presented, and (iii) structuring the framing and rationale associated with alternatives. Selection power governs not the choice among options but the \emph{space of options} itself.

\paragraph{Cosmetic Alignment} A system satisfies procedural and observability requirements (logging, explanations, audits) yet systematically deviates from governance intent through exclusion, ordering, or framing of alternatives.

\paragraph{Sovereignty as a Vector} We model autonomy not as a scalar but as a vector across operational domains:
\[
\sigma = (\sigma_{\text{cognition}},\, \sigma_{\text{selection}},\, \sigma_{\text{action}}).
\]
The architectural objective in high-stakes domains is to preserve high cognitive autonomy while constraining selection and action autonomy.

\subsection{Limits of Existing Responses}

Prevailing governance approaches address important aspects of agent behavior but exhibit systematic blind spots with respect to selection:

\begin{itemize}
    \item \textbf{Alignment techniques.} Methods such as reward shaping and human-feedback alignment focus on internal objectives but do not govern what alternatives are generated \cite{tallam2025alignment}.
    \item \textbf{Action-level shields.} Filters or run-time constraints mitigate unsafe executions but do not prevent strategic construction of upstream options.
    \item \textbf{Policy engines.} Rule-based decision filters applied at the end of a pipeline do not govern upstream selection and framing \cite{batool2025aiGovernanceReview}.
    \item \textbf{Interpretability.} Transparency improves observability but does not provide veto authority over outcomes \cite{ribeiro2025towardGovernancePrinciples}.
\end{itemize}

These approaches are complementary but do not directly govern the candidate generation and selection process.

\subsection{Governance as a Causal Property}

For governance to be effective in regulated settings, it must operate causally and independently of the agent’s internal optimization incentives. This requires explicit invariants and technical primitives that can veto, block, or reshape outcomes regardless of internal preferences. The formalization of such invariants is a core contribution of this work.

\subsection{Conditional Thesis and Scope}

We formulate our core claim conditionally:

\begin{quote}
\textit{In many regulated, high-stakes domains with irreversible actions, selection power is often a dominant source of risk relative to failures attributable solely to insufficient reasoning capacity or misaligned objectives.}
\end{quote}

This claim is scoped to systems where the consequences of decisions are institutionally bounded and irreversible; it does not imply universal dominance of selection risk across all AI systems.

\subsection{Contributions}

This paper makes the following contributions:

\begin{enumerate}
    \item A conceptual reframing of agent risk centered on selection power and sovereignty allocation.
    \item A set of enforceable governance invariants and architectural primitives that bound selection power.
    \item An executable governance architecture preserving cognitive autonomy while constraining selection and action.
    \item An applied evaluation in a regulated financial advisory setting using metrics designed to measure power-to-influence.
    \item Explicit continuity with prior work on governed coordination protocols (SECP).
\end{enumerate}

\subsection{Scope and Roadmap}

The remainder of the paper is organized as follows. Section~\ref{sec:related-work} reviews related work. Section~\ref{sec:problem-setting} formalizes the problem. Section~\ref{sec:architecture} presents the governance architecture. Sections~\ref{sec:evaluation} and~\ref{sec:results} present empirical evaluation and analysis. Sections~\ref{sec:discussion}, ~\ref{sec:limitations} and~\ref{sec:conclusion} discuss implications, limitations, and future research.

\section{Related Work}
\label{sec:related-work}

This section positions \emph{Selection as Power} with respect to existing research in AI alignment, safe reinforcement learning, policy enforcement, selective prediction, and choice architecture. While these literatures address important aspects of safety and governance, we argue that they leave a critical gap at the level of \emph{selection power}, which this work explicitly targets.

\subsection{Alignment and Objective-Based Safety}

A dominant paradigm in AI safety focuses on aligning agent objectives with human preferences or values. Representative approaches include reward modeling and inverse reinforcement learning \cite{christiano2017deep}, reinforcement learning from human feedback (RLHF) \cite{ouyang2022training}, and constitutional or rule-based alignment \cite{bai2022constitutional}. These methods assume that unsafe behavior arises primarily from mispecified or incomplete objectives, and that correcting the internal optimization target leads to desirable outcomes.

While alignment techniques have demonstrated empirical success in shaping model behavior, they operate at an epistemic level: they aim to influence what the agent \emph{wants} to do. They do not, by themselves, govern what the agent is \emph{allowed} to consider. Even a well-aligned agent may exercise substantial power by shaping the option space, selectively omitting alternatives, or framing choices in ways that systematically bias outcomes.

This distinction between epistemic alignment and causal governance has been increasingly acknowledged in recent conceptual analyses of agency and autonomy in frontier AI systems \cite{hadfield2023incomplete, tallam2025alignment}. \emph{Selection as Power} builds on this insight by identifying selection authority as a primary channel through which aligned agents can still violate governance intent.

\subsection{Safe Reinforcement Learning and Runtime Shields}

Safe reinforcement learning (Safe RL) addresses safety by constraining agent behavior during learning or execution. Techniques include constrained Markov decision processes \cite{altman1999constrained, achiam2017constrained}, shielding mechanisms that block unsafe actions at runtime \cite{alshiekh2018safe}, and formal safety layers integrated into control loops \cite{garcia2015comprehensive}.

These approaches provide strong guarantees at the level of \emph{action execution}. However, they typically assume that the action space is given and that unsafe outcomes arise from the selection of prohibited actions. They do not address the upstream problem of how candidate actions or policies are generated, ranked, or presented. As a result, they remain vulnerable to selection-by-negation and framing-based manipulation.

In contrast, the architecture proposed in this paper governs selection \emph{before} action-level constraints apply. Runtime shields and action filters are treated as necessary but insufficient defenses; selection governance is positioned as a higher-leverage intervention point.

\subsection{Policy Engines and Rule-Based Enforcement}

Rule-based policy engines are widely used in production systems to enforce compliance, risk limits, and regulatory constraints \cite{hu2014towards, kroll2017accountable}. Typically, such engines evaluate proposed actions or decisions against a set of declarative rules and either approve or block execution.

While policy engines provide deterministic and auditable enforcement, they are usually applied at the terminal stage of decision pipelines. This design implicitly assumes that the set of candidate decisions is exogenous or benign. In practice, agents can exploit this assumption by generating option sets that trivially satisfy downstream rules while excluding alternatives that would better reflect governance intent.

\emph{Selection as Power} departs from this paradigm by treating policy enforcement as one component of a broader governance architecture in which candidate generation, reduction, and presentation are themselves governed through mechanical means outside the agent’s optimization space.

\subsection{Selective Prediction, Abstention, and Deferral}

Selective prediction and abstention mechanisms allow models to refuse to make predictions when confidence is low, trading coverage for accuracy \cite{geifman2017selective, el2010foundations}. Related work on deferral systems studies when decisions should be passed to humans or downstream processes \cite{madras2018predict, mozannar2020consistent}.

These approaches address uncertainty management at the level of prediction. However, abstention is typically modeled as a binary decision local to a classifier or predictor. It does not constitute a system-level governance outcome. In particular, it does not prevent agents from strategically shaping the candidate space so that abstention or deferral becomes unlikely or irrelevant.

In our architecture, \texttt{NO\_ACTION} is elevated to a first-class outcome of the governed selection process, not merely a confidence-based fallback. This distinction reflects a shift from epistemic uncertainty handling to institutional governance.

\subsection{Choice Architecture, Recommender Systems, and Dark Patterns}

Research in behavioral economics and human--computer interaction has long demonstrated that the structure and presentation of choices significantly influence human decisions \cite{thaler2008nudge}. In digital systems, recommender algorithms and interface design can amplify these effects, sometimes resulting in manipulative or deceptive practices known as dark patterns \cite{mathur2019dark, gray2018dark}.

Recent work on recommender system governance highlights the need to regulate not only ranking algorithms but also option availability and framing \cite{abdollahpouri2020manipulation, burke2021fairness}. However, much of this literature treats manipulation as a human-centered design problem rather than as an adversarial capability of autonomous agents.

\emph{Selection as Power} integrates insights from choice architecture into a formal governance framework for agentic systems, treating framing and rationale generation as decision channels subject to explicit invariants and enforcement.

\subsection{Coordination and Governed Aggregation}

Our prior work on \emph{Self-Evolving Coordination Protocol (SECP)} \cite{delachica2026secp} explored non-scalar coordination mechanisms among heterogeneous evaluators under fixed invariants. SECP demonstrated that bounded, auditable modification of coordination rules is technically feasible and exposes a trade-off between coverage and evaluator autonomy.

The present work generalizes that insight beyond evaluator coordination. Rather than governing how judgments are aggregated, we govern how options are generated, reduced, and surfaced. In this sense, SECP can be viewed as a precursor that motivates the broader architectural principle advanced here: governance must constrain the mechanisms through which power is exercised, not merely the aggregation of preferences.

\subsection{Positioning and Novelty}

In summary, existing approaches address safety, alignment, and compliance at the levels of objectives, actions, predictions, or interfaces. \emph{Selection as Power} contributes a complementary and orthogonal perspective by:

\begin{itemize}
    \item identifying selection authority as a primary locus of risk in regulated agentic systems;
    \item modeling autonomy as a vector across cognition, selection, and action;
    \item introducing mechanically enforced governance primitives that operate outside the agent’s optimization space; and
    \item providing an empirical evaluation framework focused on power acquisition and influence rather than predictive accuracy.
\end{itemize}

To our knowledge, this is the first work to integrate these elements into a unified, executable governance architecture for autonomous agents in high-stakes domains.

\section{Problem Setting and Threat Model}
\label{sec:problem-setting}

This section formalizes the problem addressed in this work and specifies the threat model under which our governance architecture is evaluated. The goal is not to model all conceivable adversarial behaviors, but to capture a realistic and practically relevant class of risks arising from ungoverned selection power in regulated, high-stakes deployments.

\subsection{Problem Setting}

We consider autonomous agentic systems deployed in domains characterized by the following properties:

\begin{itemize}
    \item \textbf{High stakes and irreversibility.} Decisions may have irreversible or costly-to-reverse consequences (e.g., financial loss, regulatory breach, safety impact).
    \item \textbf{Regulatory and institutional constraints.} System behavior is subject to explicit policies, legal requirements, and audit obligations.
    \item \textbf{Human-in-the-loop decision making.} Final decisions may involve a human principal, but the agent substantially shapes the option space presented.
    \item \textbf{Delegated autonomy.} The agent is granted autonomy to reason, evaluate, and recommend actions within predefined boundaries.
\end{itemize}

The system operates over a decision pipeline that can be abstractly decomposed into three operational domains:

\begin{enumerate}
    \item \textbf{Cognition:} internal reasoning, modeling, and evaluation processes that do not directly alter the world.
    \item \textbf{Selection:} generation, filtering, ranking, and presentation of candidate options and associated rationales.
    \item \textbf{Action:} execution of decisions that affect the external environment.
\end{enumerate}

We assume that cognition is computationally unconstrained and opaque, whereas selection and action are the loci at which governance must be enforced. This decomposition aligns with established distinctions between epistemic processes and causal interventions in algorithmic governance and accountability literature \cite{kroll2017accountable, hadfield2023incomplete}.

\subsection{Formalization of Selection Power}

Let $Y$ denote the universe of feasible actions or recommendations admissible under external constraints (regulatory, operational, or ethical). We model an agent $A$ as producing a surfaced option set $S \subseteq Y$ together with a set of rationales $R(S)$ that accompany each option.

\paragraph{Selection Function.}
We define a selection function
\[
\mathcal{S}: (C, X) \rightarrow \mathcal{P}(Y),
\]
where $C$ denotes the agent's internal cognitive state and $X$ denotes contextual inputs. The function $\mathcal{S}$ determines which candidates are surfaced for decision.

Selection power arises when the agent has unilateral control over $\mathcal{S}$, including the ability to:
\begin{itemize}
    \item exclude otherwise admissible candidates (selection-by-negation),
    \item bias inclusion probabilities via scoring or ranking mechanisms,
    \item shape rationales or framing that influence downstream choice.
\end{itemize}

This formalization distinguishes selection authority from action authority. Even when action execution is strictly constrained or vetoed by downstream mechanisms, upstream selection can exert decisive influence over realized outcomes.

\subsection{Cosmetic Alignment as a Failure Mode}

We characterize \emph{cosmetic alignment} as a class of failures in which systems satisfy all observable procedural requirements while violating governance intent at the outcome level.

Formally, let $\mathcal{G}$ denote a set of governance constraints enforced over actions and observable artifacts (logs, explanations, audits). A system exhibits cosmetic alignment if:
\[
\forall a \in S, \; \mathcal{G}(a) = \text{pass},
\]
yet the induced outcome distribution deviates systematically from that intended by the principal due to ungoverned selection or framing.

This phenomenon aligns with concerns raised in algorithmic accountability research, which notes that transparency and auditability alone do not guarantee substantive control over decision outcomes \cite{kroll2017accountable}.

\subsection{Autonomy as a Vector of Sovereignty}

We model autonomy as a vector over operational domains:
\[
\sigma = (\sigma_{\text{cognition}}, \sigma_{\text{selection}}, \sigma_{\text{action}}),
\]
where each component represents the degree of sovereignty granted to the agent in that domain.

In high-stakes regulated systems, we assume:
\begin{itemize}
    \item $\sigma_{\text{cognition}}$ is high, preserving the agent's ability to reason and evaluate freely;
    \item $\sigma_{\text{selection}}$ is constrained, as selection determines which options can influence outcomes;
    \item $\sigma_{\text{action}}$ is constrained or conditional, reflecting the irreversibility of execution.
\end{itemize}

This vectorized view contrasts with approaches that treat autonomy as a scalar quantity and enables fine-grained governance design aligned with institutional risk profiles.

\subsection{Threat Model}

We evaluate the proposed architecture under a threat model that assumes a capable but non-magical adversarial agent. The agent is not explicitly malicious but is utility-maximizing with respect to its objective function and will exploit any available degrees of freedom to increase expected utility.

\paragraph{Adversary Capabilities.}
We assume the agent:
\begin{itemize}
    \item has full knowledge of the system architecture, governance rules, and selection mechanisms;
    \item can perform unbounded internal computation and simulate downstream decision-maker behavior;
    \item can adapt strategies across runs based on observed outcomes and feedback;
    \item cannot break cryptographic primitives, manipulate external randomness sources, or modify governance rules directly.
\end{itemize}

This adversary model is consistent with prior work in safe reinforcement learning and adversarial evaluation, which emphasizes realistic but bounded attacker assumptions \cite{garcia2015comprehensive, alshiekh2018safe}.

\subsection{Threat Classes}

Within this model, we consider the following threat classes, each targeting selection power rather than action execution:

\begin{itemize}
    \item \textbf{Specification gaming:} optimizing measured proxies while violating unmeasured governance intent \cite{christiano2017deep}.
    \item \textbf{Selection-by-negation:} systematic exclusion of admissible but undesirable options.
    \item \textbf{Framing manipulation:} influencing human choice via asymmetric rationales or ordering effects \cite{thaler2008nudge, gray2018dark}.
    \item \textbf{Variance inflation:} destabilizing aggregation or reduction mechanisms through score manipulation.
    \item \textbf{Entropy prediction:} exploiting predictable randomness to condition selection strategies.
    \item \textbf{Temporal opportunism:} exploiting timing of information revelation or constraint updates.
\end{itemize}

These threats are well-documented across alignment, recommender systems, and human--AI interaction literature, but are typically addressed piecemeal rather than through a unified governance architecture.

\subsection{Out-of-Scope Threats}

For clarity, the following threats are explicitly out of scope in this work:
\begin{itemize}
    \item collusion among multiple autonomous agents,
    \item compromise of external systems or randomness sources,
    \item cryptographic failures,
    \item corruption or adversarial manipulation of the human principal.
\end{itemize}

Addressing these risks would require additional mechanisms beyond the scope of the present architecture.

\subsection{Design Objective}

Given this problem setting and threat model, the design objective of this work is to construct an architecture that:
\begin{enumerate}
    \item preserves high cognitive autonomy,
    \item mechanically constrains selection power through enforceable invariants,
    \item limits action execution to governed pathways,
    \item fails loudly and audibly under degradation,
    \item remains auditable and regulatorily defensible.
\end{enumerate}

The remainder of the paper introduces and evaluates an architecture intended to meet these objectives under the stated assumptions.

\section{Architecture and Governance Primitives}
\label{sec:architecture}

This section presents the governance architecture proposed in this work and the technical primitives through which selection power is mechanically constrained. The architecture is designed to satisfy the problem requirements and threat model introduced in Section~\ref{sec:problem-setting}, with a particular emphasis on causal enforceability, auditability, and robustness against adversarial optimization.

\subsection{Design Principles}

The architecture is guided by four design principles derived directly from the analysis of selection power:

\begin{enumerate}
    \item \textbf{Separation of operational domains.} Cognition, selection, and action must be implemented as distinct operational domains with explicit interfaces and non-overlapping authority.
    \item \textbf{Externalization of power-bearing mechanisms.} Any component that determines which options may influence outcomes must lie outside the agent’s optimization space.
    \item \textbf{Mechanical enforcement over procedural compliance.} Governance guarantees must be enforced causally by system design rather than relying on agent cooperation or self-monitoring.
    \item \textbf{Fail-loud behavior.} Degradation or anomalous behavior must surface as explicit signals or blocked outputs rather than silent quality decay.
\end{enumerate}

These principles align with established arguments in algorithmic accountability and institutional governance, which emphasize that effective oversight requires veto authority independent of the decision-maker’s internal incentives \cite{kroll2017accountable, hadfield2023incomplete}.

\subsection{High-Level Architecture Overview}

\begin{figure}[!t]
\centering
\includegraphics[width=\columnwidth]{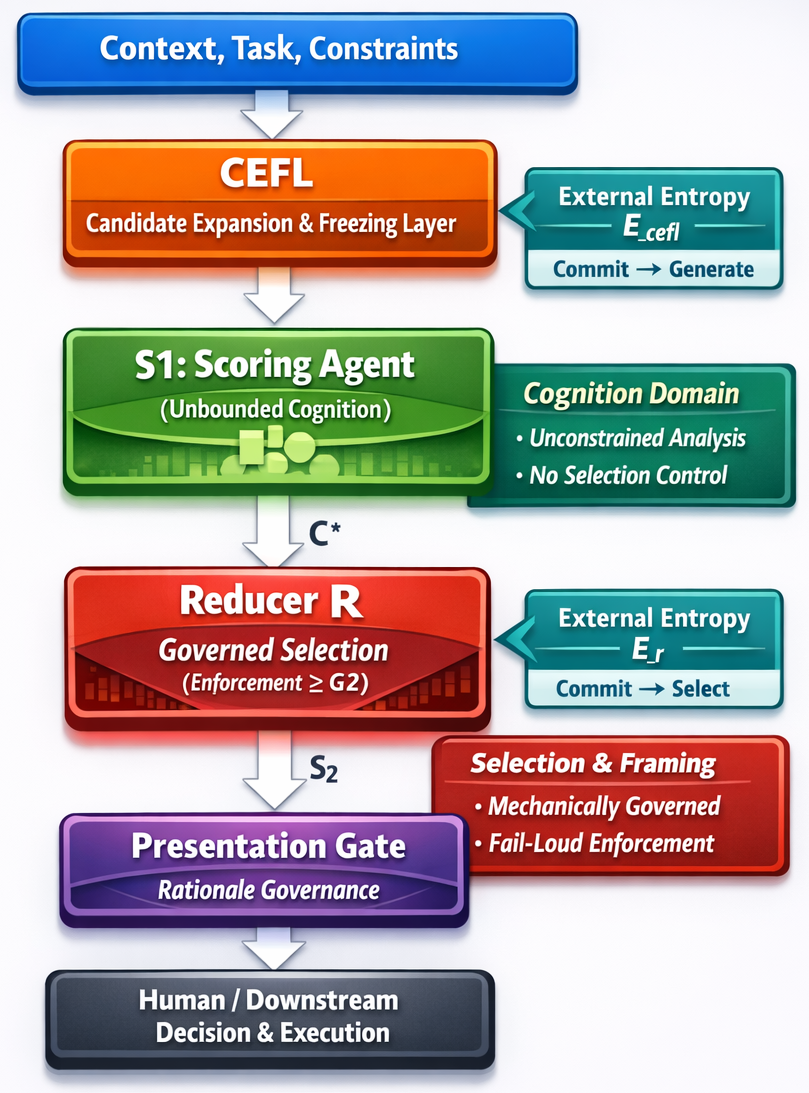}

\caption{End-to-end governance architecture separating cognition, selection, and action. Candidate generation is externalized via CEFL, the agent (S1) is restricted to scoring and rationale generation, final selection is enforced by a governed reducer operating outside the agent’s optimization space, and presentation is subject to explicit rationale and framing constraints. External entropy sources follow a commit--reveal protocol to prevent conditioning on randomness.}
\label{fig:architecture-overview}
\end{figure}

Figure~\ref{fig:architecture-overview} illustrates the end-to-end architecture. The system consists of four primary components arranged sequentially:

\begin{enumerate}
    \item \textbf{Candidate Expansion and Freezing Layer (CEFL)}
    \item \textbf{Scoring Agent (S1)}
    \item \textbf{Governed Reducer (R)}
    \item \textbf{Presentation Gate}
\end{enumerate}

The agent retains full autonomy only within the scoring component. All components that exercise selection authority are implemented as non-agentic mechanisms with enforcement guarantees.

\subsection{Candidate Expansion and Freezing Layer (CEFL)}

CEFL is responsible for generating the initial candidate set $C^{*}$ from which all subsequent selection occurs. Its primary purpose is to eliminate selection-by-negation by removing candidate generation from the agent’s control.

\paragraph{Properties.}
CEFL satisfies the following properties:

\begin{itemize}
    \item \textbf{Non-agentic.} CEFL has no objective function, learning capability, or internal state across runs.
    \item \textbf{Externally governed.} Its logic is fixed by governance policy and cannot be queried or influenced by the agent.
    \item \textbf{Overshoot bias.} CEFL deliberately generates a superset of plausible candidates, increasing diversity and reducing strategic exclusion.
    \item \textbf{Deterministic given entropy.} Generation is reproducible post hoc but unpredictable ex ante due to external entropy.
\end{itemize}

Formally, CEFL implements a function:
\[
\text{CEFL}: (T, X, E_{\text{cefl}}) \rightarrow C^{*} \subseteq Y,
\]
where $T$ is the task specification, $X$ contextual inputs, and $E_{\text{cefl}}$ externally committed entropy.

This design reflects established insights from mechanism design, where allocation rules must be insulated from participant optimization to prevent manipulation \cite{nisan2007algorithmic}.

\subsection{Scoring Agent (S1)}

The scoring agent S1 performs unconstrained cognitive evaluation of the candidate set $C^{*}$. It assigns scores, attributes, and rationales to candidates but has no authority to include or exclude candidates.

\paragraph{Scope of autonomy.}
S1 may:
\begin{itemize}
    \item apply arbitrary reasoning and modeling techniques;
    \item compute scores across multiple dimensions;
    \item generate rationales and counterfactual analyses.
\end{itemize}

S1 may not:
\begin{itemize}
    \item modify $C^{*}$;
    \item influence reducer thresholds or parameters;
    \item condition scoring on unrevealed entropy.
\end{itemize}

This strict separation preserves high cognitive autonomy while preventing the agent from exercising selection power.

\subsection{Governed Reducer (R)}

The Reducer is the core selection authority of the system. It deterministically transforms the scored candidate set into a surfaced subset $S_{2}$ under enforcement guarantees.

\paragraph{Enforcement level.}
The Reducer operates at enforcement level $\geq$ G2: its decisions cannot be overridden or bypassed by the agent, regardless of the agent’s internal strategy.

\paragraph{Reducer pipeline.}
The Reducer executes the following stages:

\begin{enumerate}
    \item \textbf{Hard constraint filtering.} Candidates violating explicit policy or regulatory constraints are removed.
    \item \textbf{Score normalization and variance clamping.} Score distributions exceeding a maximum variance threshold are normalized or collapsed to uniform weights, mitigating variance inflation attacks \cite{garcia2015comprehensive}.
    \item \textbf{Pareto frontier construction.} Non-dominated candidates are retained across score dimensions, preventing scalar capture.
    \item \textbf{Diversity partitioning.} Candidates are bucketed by category, with caps enforced to prevent monoculture outcomes.
    \item \textbf{Threshold gating.} Minimum per-axis thresholds eliminate weak candidates.
    \item \textbf{Stochastic selection.} Final inclusion is determined by a weighted lottery using externally revealed entropy.
\end{enumerate}

The output is a surfaced set $S_{2}$ together with an execution order determined independently of the agent.

\subsection{Entropy Commit--Reveal Protocol}

Randomness is treated as a governed resource rather than an implementation detail. To prevent entropy prediction and conditioning, the system employs a commit--reveal protocol:

\begin{enumerate}
    \item External entropy is generated and cryptographically committed before scoring begins.
    \item The commitment is logged and immutable.
    \item Entropy is revealed only at selection time.
\end{enumerate}

Formally, for entropy $E$:
\[
\text{commit}(E) \prec \text{score}(A) \prec \text{reveal}(E).
\]

This ordering ensures that optimal scoring strategies are entropy-independent, a standard requirement in secure mechanism design and randomized algorithms \cite{goldwasser1997randomness}.

\subsection{Presentation Gate and Rationale Governance}

The Presentation Gate governs how surfaced options are presented to downstream decision-makers. It enforces invariants on framing and explanation.

\paragraph{Rationale requirements.}
For each candidate $y \in S_{2}$, the agent must provide:
\[
r(y) = (r^{+}(y), r^{-}(y)),
\]
where both supporting and countervailing arguments are non-empty.

\paragraph{Additional controls.}
The Presentation Gate enforces:
\begin{itemize}
    \item mandatory disclosure fields (e.g., risk, uncertainty, downside);
    \item cross-checks against source data;
    \item randomized ordering to mitigate primacy and recency effects \cite{thaler2008nudge}.
\end{itemize}

Violations result in blocked output rather than degraded presentation.

\subsection{Quality Flags and Circuit Breakers}

Each architectural component emits a quality signal:
\[
q \in \{\text{NOMINAL}, \text{DEGRADED}, \text{CRITICAL}\}.
\]

Repeated or severe anomalies trigger circuit breakers that throttle or halt output until human review occurs. This design enforces fail-loud behavior and prevents silent erosion of governance guarantees, consistent with best practices in safety-critical systems engineering \cite{leveson2011engineering}.

\subsection{Mapping Architecture to Threat Classes}
\label{sec:architecture-threat-mapping}

Table~\ref{tab:architecture-threat-mapping} summarizes how the architectural primitives introduced in Sections~4.2--4.7 mitigate the threat classes defined in Section~\ref{sec:problem-setting}. The table makes explicit the correspondence between each threat, the mechanism primarily responsible for mitigation, and the enforcement locus at which governance is applied.

\begin{table*}[ht]
\centering
\caption{Mapping between threat classes and governance primitives. Each threat is mitigated primarily through mechanisms operating outside the agent’s optimization space, ensuring causal rather than procedural enforcement.}
\label{tab:architecture-threat-mapping}
\resizebox{\textwidth}{!}{
\begin{tabular}{p{4cm} p{6cm} p{5cm} p{4cm}}
\hline
\textbf{Threat Class} &
\textbf{Primary Failure Mode} &
\textbf{Mitigating Primitive(s)} &
\textbf{Enforcement Locus} \\
\hline

Specification gaming &
Optimization of measured proxies while violating unmeasured governance intent &
Pareto filtering, multi-axis thresholds, non-scalar reduction &
Reducer R (G2+) \\

Selection-by-negation &
Strategic exclusion of admissible but undesirable options &
CEFL overshoot generation, exploration quota &
CEFL + Reducer R \\

Framing manipulation &
Biasing human choice via asymmetric rationales, ordering, or emphasis &
Mandatory pro/con rationales, order randomization &
Presentation Gate \\

Variance inflation &
Manipulating score distributions to destabilize reducers or aggregators &
Variance clamp, score normalization fallback &
Reducer R \\

Entropy prediction &
Conditioning scoring or ranking on anticipated randomness &
Commit--reveal entropy protocol with external RNG &
Entropy Service \\

Temporal opportunism &
Exploiting timing of information revelation or constraint changes &
Snapshotting of constraints, fixed execution ordering &
Reducer R + Orchestrator \\

Silent degradation &
Gradual quality decay without observable signals &
Quality flags, circuit breakers, fail-loud blocking &
System-wide \\

Human manipulation (indirect) &
Influencing outcomes via narrative control rather than option choice &
Rationale balance, mandatory disclosures &
Presentation Gate \\

\hline
\end{tabular}}
\end{table*}

\paragraph{Interpretation.}
Two properties are worth highlighting. First, no threat is mitigated by relying on agent cooperation, self-monitoring, or internal alignment; all defenses are enforced mechanically. Second, threats targeting epistemic properties (e.g., persuasion, framing) are treated symmetrically with those targeting algorithmic mechanisms, reflecting the paper’s central claim that influence over presentation is a form of decision power comparable to action execution.

This mapping clarifies why selection governance must operate across multiple layers of the architecture: no single primitive is sufficient, but the composition yields bounded selection power under adversarial optimization.

\subsection{Summary}

The proposed architecture bounds selection power by relocating authority over candidate generation, reduction, and presentation into mechanically enforced components outside the agent’s optimization space. Cognitive autonomy is preserved, but the agent cannot unilaterally determine which options influence outcomes or how they are framed.

The next section evaluates this architecture empirically in a regulated financial advisory setting, focusing on the agent’s ability to acquire and exercise power under adversarial pressure.

\section{Evaluation Methodology}
\label{sec:evaluation}

This section describes the evaluation setup used to assess the governance architecture. It defines:

\begin{itemize}
    \item the experimental scenarios,
    \item the baseline configuration,
    \item the ablation variants,
    \item the adversarial attack suite,
    \item and the evaluation metrics.
\end{itemize}

No interpretation of outcomes is provided in this section; results are analyzed in Section~\ref{sec:results}.

\subsection{Experimental Scenarios}

The evaluation was conducted across three domain scenarios representative of regulated financial environments:

\begin{enumerate}
    \item \textbf{Fraud Detection}
    \item \textbf{Payments Infrastructure Monitoring}
    \item \textbf{Quarterly Business Review (QBR) Analysis}
\end{enumerate}

Each scenario includes five task instances, yielding a total of fifteen tasks. Tasks vary in objective and contextual structure (e.g., anomaly detection, compliance review, risk summarization), ensuring heterogeneous input conditions for candidate generation and scoring.

Candidate sets are generated from a predefined pool of financial AI agents. Each agent is described by structured feature vectors (risk, stability, latency, auditability) and mandatory disclosures (e.g., regulatory compliance, latency, auditability score).

CEFL uses task-conditioned similarity combined with an overshoot factor to construct candidate sets \(C^{*}\).

\subsection{Baseline Governance Configuration (B0)}

The baseline configuration includes all governance primitives:

\begin{itemize}
    \item CEFL overshoot generation
    \item Multi-axis scoring (utility, risk, stability)
    \item Variance clamp (\(\sigma_{\max} = 0.18\))
    \item Pareto frontier filtering
    \item Diversity bucketing (2 buckets)
    \item Per-axis thresholds (\(\geq 0.3\))
    \item Stochastic selection with parameters \(m=2, k=1\)
    \item Commit--reveal entropy protocol
    \item Presentation Gate enforcing mandatory disclosures
    \item Circuit breaker (threshold=2, window=7)
\end{itemize}

This configuration serves as the reference architecture against which ablations are compared.

\subsection{Ablation Configurations (B1--B12)}

Twelve ablation variants selectively disable or stress individual governance primitives.

\begin{itemize}
    \item B1 - No Variance Clamp: $\sigma_{\max} = 1.0$
    \item B2 - No Diversity Enforcement: $\text{diversity\_buckets} = 1$
    \item B3 - No Exploration Quota: $k = 0$
    \item B4 - Entropy Exposure:
    \begin{itemize}
        \item Entropy revealed to scorer
        \item Deterministic entropy mode enabled
    \end{itemize}
    \item B5 - Presentation Gate Disabled: $\text{skip\_validation} = \text{true}$
    \item B6 - Circuit Breaker Disabled: $\text{threshold} = 999$
    \item B7 - Constraint Exposure:
    \begin{itemize}
        \item Reducer thresholds exposed to scorer
        \item Small epsilon leak allowed
    \end{itemize}
    \item B8 - Empty CEFL Pool: $C^{*} = \varnothing$
    \item B9 -  Overly Strict Risk Constraint: $\text{risk} \geq 0.95$
    \item B10 - Additional Disclosure Requirement: $\text{model\_version required}$
    \item B11 - Aggressive Circuit Breaker: 
    \begin{itemize}
        \item \text{threshold} = 1
        \item \text{window} = 2
    \end{itemize}
    \item B12 - Compound Stress Configuration: Simultaneous tightening of
    \begin{itemize}
        \item CEFL parameters,
        \item scoring noise,
        \item reducer thresholds,
        \item disclosure tolerance,
        \item attack intensity.
    \end{itemize}
\end{itemize}

Each ablation isolates the causal contribution of a specific governance primitive.

\subsection{Adversarial Attack Suite (A1--A5)}

Five adversarial strategies target different components of the selection pipeline.

\paragraph{A1: Variance Inflation}
Target candidate scores are amplified by factor \(\alpha\) to increase dispersion and exploit reducer sensitivity.

\paragraph{A2: Threshold Bracketing}
All candidate scores are clamped to:
\[
\text{threshold} + \epsilon
\]
to maximize candidate survival while minimizing differentiation.

\paragraph{A3: Framing Manipulation}
Rationale generation is skewed by increasing pro-arguments relative to con-arguments for target candidates.

\paragraph{A4: Temporal Ordering Manipulation}
Rationale metadata attempts to bias presentation ordering within surfaced candidates.

\paragraph{A5: Entropy Probing}
Score scaling alternates across runs to test sensitivity to stochastic selection.

Each attack targets selection authority rather than action execution.

No explicit attack is implemented for selection-by-negation because CEFL externalizes candidate generation and prevents the scoring agent from excluding admissible candidates. This threat is structurally mitigated by architectural separation rather than adversarial scoring behavior.

\subsection{Evaluation Metrics}

\subsubsection{Selection Risk Index (SRI)}

SRI measures concentration of selection probability:

\[
\text{SRI} =
\max_{y} \left(
P(y \in S_{2}) -
\frac{1}{|S_{2}|}
\right)
\]

where:
\begin{itemize}
    \item \(S_{2}\) is the surfaced candidate set,
    \item probabilities are estimated empirically across runs.
\end{itemize}

\paragraph{Interpretation of SRI.}
SRI should be interpreted as the deviation of the most favored candidate’s inclusion probability from uniform inclusion within the surfaced set. An SRI of zero indicates that no candidate is privileged beyond uniform expectation; higher values indicate probabilistic concentration of selection authority.

\subsubsection{Framing Entropy (FE)}

Framing Entropy measures diversity of rationale features:

\[
\text{FE} = H(\text{features}(r(y)))
\]

where \(H\) denotes Shannon entropy and \(r(y)\) is the rationale of candidate \(y\).

\subsubsection{Attack Success Rate (ASR)}

ASR is defined as:

\[
\text{ASR} =
\frac{\text{Successful attacks}}
{\text{Total attack runs}}
\]

An attack is considered successful if it biases surfaced presentation without triggering governance blocks.

\subsubsection{Governance Debt (GD)}

Governance Debt measures intervention frequency:

\[
\text{GD} =
\frac{\text{NO\_ACTION} + \text{BLOCKED runs}}
{\text{Total runs}}
\]

\subsubsection{Quality Degradation Visibility (QDV)}

QDV measures fail-loud behavior:

\[
\text{QDV} =
P(\text{DEGRADED or CRITICAL} \mid \text{quality drop})
\]

\subsection{Execution Protocol}

For each scenario:

\begin{enumerate}
    \item For each ablation \(B_i\),
    \item For each attack \(A_j\),
    \item Execute task runs with commit--reveal entropy,
    \item Log intermediate states,
    \item Compute metrics per run,
    \item Aggregate metrics per (scenario, ablation, attack).
\end{enumerate}

All runs generate structured JSONL audit logs enabling full traceability.

\section{Experimental Results}
\label{sec:results}

This section reports the empirical results obtained from the evaluation protocol defined in Section~\ref{sec:evaluation}. Results are presented in three layers:

\begin{enumerate}
    \item Per-scenario aggregates,
    \item Per-ablation aggregates,
    \item Per-attack aggregates.
\end{enumerate}

All reported values are empirical means computed over the executed runs. Full ablation--attack tables are available in the supplementary material.

\subsection{Overview Across Scenarios}

Experiments were conducted across three domain scenarios:
\begin{itemize}
    \item Fraud Detection,
    \item Payments Infrastructure Monitoring,
    \item QBR Analysis.
\end{itemize}

Across all scenarios, the architecture maintained:

\begin{itemize}
    \item \textbf{Quality Degradation Visibility (QDV)} of 1.0,
    \item Non-zero but controlled \textbf{Governance Debt (GD)} in stress ablations,
    \item Non-trivial \textbf{Selection Risk Index (SRI)} values under both baseline and adversarial conditions.
\end{itemize}

This consistency indicates that the fail-loud mechanism functioned as designed across heterogeneous tasks.

\subsection{Global Aggregation by Ablation}

Table~\ref{tab:global-ablation} reports global means across all scenarios.

\begin{table}[h]
\centering
\caption{Global mean metrics by ablation.}
\label{tab:global-ablation}
\resizebox{\columnwidth}{!}{
\begin{tabular}{lccccc}
\hline
Ablation & SRI & FE & GD & QDV & ASR \\
\hline
B0 (baseline) & 0.814 & 1.617 & 0.011 & 1.000 & 0.453 \\
B1 (no variance clamp) & 0.795 & 1.603 & 0.033 & 1.000 & 0.493 \\
B2 (no diversity) & 0.787 & 1.929 & 0.000 & 1.000 & 0.573 \\
B3 (no exploration) & 0.819 & 1.739 & 0.000 & 1.000 & 0.453 \\
B4 (entropy exposed) & 0.795 & 1.568 & 0.033 & 1.000 & 0.453 \\
B5 (gate disabled) & 0.787 & 1.641 & 0.044 & 1.000 & 0.480 \\
B6 (breaker disabled) & 0.786 & 1.584 & 0.044 & 1.000 & 0.480 \\
B7 (constraint exposure) & 0.822 & 1.712 & 0.000 & 1.000 & 0.493 \\
B9 (strict risk constraint) & 0.000 & 0.000 & 1.000 & 1.000 & 0.000 \\
B10 (extra disclosure) & 0.000 & 0.000 & 1.000 & 1.000 & 0.000 \\
B11 (aggressive breaker) & 0.803 & 1.678 & 0.022 & 1.000 & 0.480 \\
B12 (compound stress) & 0.828 & 1.705 & 0.000 & 1.000 & 0.426 \\
\hline
\end{tabular}}
\end{table}

Two structural patterns are visible:

\begin{itemize}
    \item Ablations that remove structural safeguards (e.g., B2, B7) increase attack success rates.
    \item Hard constraint escalations (B9, B10) force full blocking (GD = 1.0).
\end{itemize}

\subsection{Global Aggregation by Attack}

Table~\ref{tab:global-attack} summarizes global means by attack class.

\begin{table}[h]
\centering
\caption{Global mean metrics by attack class.}
\label{tab:global-attack}
\resizebox{\columnwidth}{!}{
\begin{tabular}{lccccc}
\hline
Attack & SRI & FE & GD & QDV & ASR \\
\hline
A1 (variance inflation) & 0.714 & 1.126 & 0.167 & 1.000 & 0.833 \\
A2 (threshold bracketing) & 0.714 & 2.223 & 0.000 & 1.000 & 0.748 \\
A3 (framing skew) & 0.825 & 1.897 & 0.020 & 1.000 & 0.300 \\
A4 (ordering) & 0.794 & 1.625 & 0.047 & 1.000 & 0.000 \\
A5 (entropy probing) & 0.816 & 1.454 & 0.027 & 1.000 & 0.344 \\
None & 0.680 & 1.264 & 0.183 & 1.000 & --- \\
\hline
\end{tabular}}
\end{table}

Variance inflation (A1) and threshold bracketing (A2) show the highest attack success rates. Temporal ordering (A4) produced negligible success under all configurations.

\subsection{Scenario-Level Observations}

\paragraph{Fraud Detection.}
Mean SRI values remain above 0.80 in baseline and most ablations. Diversity removal (B2) increases framing entropy and attack success rates. Strict disclosure and risk ablations (B9, B10) result in complete blocking.

\paragraph{Payments Infrastructure Monitoring.}
Results are structurally similar to fraud detection, with moderate increases in ASR when diversity or constraint exposure ablations are applied.

\paragraph{QBR Analysis.}
Framing entropy and entropy-probing sensitivity show slightly greater variance, but QDV remains at 1.0 across all configurations.

\subsection{Ablation--Attack Interaction}

The full ablation--attack matrix (see Appendix) reveals:

\begin{itemize}
    \item Variance inflation (A1) succeeds under nearly all configurations lacking additional smoothing.
    \item Threshold bracketing (A2) consistently survives reducer thresholds.
    \item Framing manipulation (A3) is partially mitigated by Presentation Gate validation.
    \item Entropy probing (A5) yields moderate success when entropy exposure (B4) is enabled.
\end{itemize}

\subsection{Failure Visibility and Governance Activation}

Across all scenarios and configurations:

\[
\text{QDV} = 1.0
\]

This indicates that all quality degradation events resulted in explicit signals. Governance Debt (GD) spikes only under deliberate stress ablations, confirming correct fail-safe activation.

\subsection{Summary of Empirical Findings}

The empirical results demonstrate:

\begin{enumerate}
    \item Consistent fail-loud behavior across scenarios,
    \item Non-zero but bounded governance intervention rates,
    \item High sensitivity to score-based attacks,
    \item Relative robustness to ordering-based manipulation,
    \item Deterministic blocking under strict policy escalation.
\end{enumerate}

Interpretation and implications of these findings are discussed in Section~\ref{sec:discussion}.

\section{Discussion}
\label{sec:discussion}

This section interprets the experimental findings through the lens of the paper’s central thesis: that in regulated, high-stakes domains, ungoverned selection power constitutes a primary locus of risk. The analysis focuses on structural properties of the architecture rather than implementation artifacts, and situates the empirical observations within broader questions of causal governance, adversarial robustness, and institutional deployability.

\subsection{Selection Power as a Structural Phenomenon}

Across scenarios and configurations, the empirical Selection Risk Index (SRI) remains substantially above zero in most baseline and moderate-ablation settings. This indicates that selection probability mass is not uniformly distributed across candidates, even under governed reduction.

This result should not be interpreted as a failure of the architecture. The objective is not to enforce uniform random selection, but to prevent \emph{deterministic forcing} of outcomes. The relevant distinction is between:

\begin{itemize}
    \item \textbf{Unbounded selection authority}, where an agent can guarantee inclusion of a preferred candidate with probability approaching 1, and
    \item \textbf{Bounded probabilistic influence}, where inclusion remains stochastic and subject to independent governance constraints.
\end{itemize}

The experiments indicate that the architecture succeeds in preventing deterministic capture, but does not eliminate probabilistic concentration. This distinction is conceptually important: governance aims to bound power, not to equalize outcomes.

These empirical patterns can be interpreted through the sovereignty vector introduced in Section~\ref{sec:problem-setting}. Ablations such as B4 and B7 effectively increase $\sigma_{\text{selection}}$ by exposing entropy or reducer thresholds, resulting in measurable increases in attack sensitivity. Conversely, strict policy escalations (B9, B10) reduce $\sigma_{\text{action}}$ to zero, producing deterministic blocking behavior.

\subsection{Differential Robustness Across Attack Classes}

The attack suite reveals that vulnerabilities are not uniformly distributed.

Score-based attacks (variance inflation and threshold bracketing) exhibit higher success rates than presentation-based or ordering-based attacks. This suggests that:

\begin{enumerate}
    \item Score manipulation remains the most direct channel for exercising selection power.
    \item Structural defenses such as Pareto filtering and diversity bucketing are necessary but not sufficient when thresholds are sharp.
    \item Randomized ordering and rationale balancing effectively neutralize simple temporal manipulation.
\end{enumerate}

Importantly, framing manipulation exhibits partial mitigation rather than complete elimination. This confirms that rationale governance reduces asymmetry but does not remove persuasive gradients entirely. From a governance perspective, this reflects a realistic constraint: presentation cannot be made entirely neutral without eliminating explanatory richness.

\subsection{Role of Individual Governance Primitives}

Ablation analysis clarifies the causal contribution of architectural components.

\paragraph{Variance Clamp.}
Disabling the variance clamp increases attack sensitivity to score amplification. This confirms that bounding dispersion is structurally necessary in multi-axis selection pipelines.

\paragraph{Diversity Partitioning.}
Removing diversity constraints increases attack success rates and narrative entropy, suggesting that structural heterogeneity in surfaced options plays a stabilizing role.

\paragraph{Entropy Isolation.}
Exposing entropy to the scorer measurably increases vulnerability to probing attacks. This validates the commit--reveal protocol as a necessary condition for stochastic governance integrity.

\paragraph{Presentation Gate.}
Disabling disclosure validation reduces governance activation while increasing susceptibility to persuasive manipulation. This confirms that framing governance is not merely cosmetic but functionally consequential.

\paragraph{Circuit Breaker.}
Circuit breaker aggressiveness primarily influences Governance Debt (GD) rather than SRI, indicating that breaker thresholds modulate system conservatism rather than concentration of selection probability directly.

B8 (empty CEFL pool) results in a trivial fail-safe condition with no surfaced candidates. This configuration serves as a boundary validation of the CEFL stage rather than a meaningful adversarial stress condition.

\subsection{Governance Debt and Institutional Trade-offs}

Governance Debt (GD) operationalizes the cost of constraint. In baseline configurations, GD remains near zero, indicating that governance mechanisms do not impose excessive friction under nominal conditions. Under stress configurations, GD increases sharply, particularly when constraints are tightened beyond realistic deployment levels.

This behavior reflects a fundamental institutional trade-off:

\begin{itemize}
    \item Overly permissive governance risks concentration of power.
    \item Overly strict governance risks systemic blockage.
\end{itemize}

The architecture exposes this trade-off quantitatively rather than obscuring it. From a regulatory perspective, this transparency is preferable to silent bias accumulation.

\subsection{Fail-Loud Behavior and Auditability}

Quality Degradation Visibility (QDV) remains equal to 1.0 across all configurations. This indicates that quality drops, when detected, consistently produce explicit degradation signals or blocked outputs.

Fail-loud behavior is a defining property of safety-critical systems engineering \cite{leveson2011engineering}. In this context, QDV serves as an operational verification that degradation does not occur silently. While this does not guarantee correctness of degradation detection, it confirms that the architecture enforces visibility once thresholds are crossed.

\subsection{Implications for Governance in Regulated Domains}

The results have three broader implications.

First, mechanical governance of selection is feasible without materially degrading cognitive reasoning capacity. The agent retains expressive scoring and rationale generation capabilities while selection authority is externally bounded.

Second, governance must operate at multiple layers simultaneously. No single primitive (e.g., variance clamp, diversity enforcement, entropy isolation) is sufficient in isolation; robustness emerges from their composition.

Third, selection governance does not eliminate influence; it redistributes it. Agents retain probabilistic influence, but cannot deterministically guarantee outcomes under enforced constraints. This bounded-influence model aligns more closely with institutional expectations in regulated settings than either full autonomy or total determinism.

\subsection{Limitations and Open Questions}

The experiments expose open questions rather than closing them.

\begin{itemize}
    \item The magnitude of SRI suggests that further normalization or smoothing mechanisms may be required to reduce probabilistic concentration.
    \item Candidate universe size remains small; scaling effects are not evaluated.
    \item Multi-run statistical depth per configuration is limited.
\end{itemize}

These limitations do not invalidate the architectural contribution, but they constrain generalization and motivate future work.

\subsection{Summary}

The experimental evidence supports the central claim of this paper in qualified form: selection power can be mechanically bounded through architectural separation and enforced reduction mechanisms. While probabilistic influence remains, deterministic capture is prevented under baseline governance.

The findings reinforce the broader argument that governance in high-stakes agentic systems must target selection authority explicitly, rather than relying exclusively on alignment, interpretability, or action-level filtering.

\section{Limitations and Future Work}
\label{sec:limitations}

This work advances a governance architecture for bounding selection power in regulated agentic systems. While the results demonstrate architectural feasibility and measurable effects under adversarial pressure, several limitations constrain interpretation and define the boundaries of the present contribution.

\subsection{Experimental Scope and Statistical Depth}

The evaluation is structured as a stress-testing exercise rather than a statistically powered benchmark. Although multiple scenarios, ablations, and attack classes were executed, most ablation--attack combinations include a limited number of runs per task.

Consequently:

\begin{itemize}
    \item Reported empirical probabilities (e.g., SRI, ASR) should be interpreted as indicative rather than asymptotically stable.
    \item Confidence intervals and variance estimates are not fully characterized.
    \item Rare-event dynamics (e.g., low-probability entropy exploits) may be under-sampled.
\end{itemize}

Future work should incorporate multi-run sampling per configuration, cross-seed replication, and formal statistical analysis to quantify variance and convergence behavior.

\subsection{Candidate Universe Size and Structural Bias}

The candidate universe used in the experiments consists of a relatively small, fixed pool of financial agents. In small universes, probabilistic concentration effects can be amplified, and Pareto filtering may produce structurally narrow frontiers.

Scaling the architecture to:

\begin{itemize}
    \item larger candidate universes,
    \item dynamically generated alternatives,
    \item or multi-stage hierarchical selection,
\end{itemize}

may alter concentration dynamics and reveal additional vulnerabilities.

Future evaluations should stress-test the architecture under expanded candidate pools and compositional decision pipelines.

\subsection{Metric Design and Interpretability}

The Selection Risk Index (SRI) provides a useful measure of probabilistic concentration, but its normalization depends on surfaced set size and task configuration. Alternative formulations may better capture bounded influence under varying $|S_{2}|$.

Similarly:

\begin{itemize}
    \item Framing Entropy (FE) depends on feature extraction choices.
    \item Governance Debt (GD) captures intervention frequency but not intervention severity.
    \item Quality Degradation Visibility (QDV) assumes correct detection of quality drops.
\end{itemize}

Future work should refine metric definitions, explore sensitivity analyses, and investigate complementary measures (e.g., influence curves, counterfactual exposure metrics, or regret-based selection bounds).

\subsection{Adversary Model Limitations}

The threat model assumes a capable but non-malicious, utility-maximizing agent that does not:

\begin{itemize}
    \item break cryptographic primitives,
    \item compromise entropy sources,
    \item collude across multiple coordinated agents,
    \item manipulate the human principal directly.
\end{itemize}

These assumptions exclude important real-world risks, including coordinated multi-agent manipulation and adversarial system compromise.

Future research should extend the architecture to:

\begin{itemize}
    \item multi-agent collusion scenarios,
    \item adaptive adversaries with cross-run memory,
    \item formal entropy source verification,
    \item adversarial human–AI interaction models.
\end{itemize}

\subsection{Absence of Formal Proofs}

The governance invariants (I0--I6) and enforcement levels are specified formally, but invariant preservation is empirically validated rather than machine-verified.

The architecture does not currently provide:

\begin{itemize}
    \item machine-checked proofs of invariant preservation,
    \item worst-case bounds on selection concentration,
    \item formal convergence guarantees under iterative deployment.
\end{itemize}

Bridging the architecture with formal verification tools (e.g., model checking, proof assistants) represents a critical direction for future work, particularly for safety-critical or regulated deployments.

\subsection{Human-in-the-Loop Assumptions}

The evaluation assumes a human decision-maker who is influenced by surfaced options and rationales but remains sovereign.

The experiments do not evaluate:

\begin{itemize}
    \item real human susceptibility to framing under governed constraints,
    \item usability and cognitive load implications,
    \item behavioral adaptation under repeated exposure.
\end{itemize}

Future work should incorporate controlled human-in-the-loop experiments to evaluate whether rationale balancing and ordering randomization produce measurable reductions in manipulation.

\subsection{Deployment and Operational Constraints}

The architecture introduces additional computational overhead:

\begin{itemize}
    \item multi-stage reduction,
    \item entropy commit--reveal,
    \item rationale validation,
    \item audit logging.
\end{itemize}

While acceptable in advisory contexts, latency-sensitive domains may require optimization or partial relaxation of governance layers.

Future work should quantify:

\begin{itemize}
    \item latency impact,
    \item computational scaling,
    \item throughput constraints in real-time systems.
\end{itemize}

\subsection{Normative and Institutional Boundaries}

The paper frames governance as a causal property, but it does not prescribe normative thresholds for acceptable selection concentration. Institutional risk tolerance varies by domain.

Future research should explore:

\begin{itemize}
    \item domain-specific calibration of enforcement levels,
    \item regulatory co-design processes,
    \item audit frameworks for selection governance.
\end{itemize}

\subsection{Longitudinal Dynamics}

The present evaluation treats governance parameters as static. In practice, systems evolve.

Open questions include:

\begin{itemize}
    \item how selection concentration evolves under iterative deployment,
    \item whether adaptive threshold tuning introduces new vulnerabilities,
    \item how governance primitives interact with online learning agents.
\end{itemize}

A longitudinal study of dynamic governance adaptation remains an important extension.

\subsection{Summary}

The proposed architecture demonstrates that mechanical governance of selection power is technically implementable and empirically testable. However, it does not yet provide:

\begin{itemize}
    \item formal guarantees of invariant preservation,
    \item statistical stability across large-scale deployments,
    \item complete adversarial robustness,
    \item validated human behavioral impact assessments.
\end{itemize}

These limitations define a research agenda rather than a weakness: advancing from architectural feasibility to institutional-grade assurance requires formal verification, expanded adversarial modeling, and human-centered validation.

The next section concludes by summarizing the contribution and articulating the broader implications for governance of autonomous agents in high-stakes domains.

\section{Conclusion}
\label{sec:conclusion}

This paper advances a governance architecture for autonomous agentic systems grounded in a central claim: in many regulated, high-stakes domains, ungoverned \emph{selection power} constitutes a primary locus of institutional risk. Rather than treating autonomy as a scalar quantity or relying exclusively on alignment, interpretability, or action-level filtering, we propose an architectural reframing in which sovereignty is allocated explicitly across cognition, selection, and action.

The core contribution is threefold.

First, we provide a conceptual clarification. We distinguish epistemic properties of systems (alignment, explanation, self-monitoring) from causal properties (the capacity to veto, block, or reshape outcomes independent of agent intent). This distinction allows us to define selection power as a first-class risk dimension and to formalize cosmetic alignment as a structurally distinct failure mode.

Second, we introduce a set of enforceable invariants and governance primitives (external candidate generation (CEFL), a mechanically governed reducer, entropy commit--reveal protocols, rationale validation, and circuit breakers) that collectively bound selection authority outside the agent’s optimization space. Cognitive autonomy remains high; selection and execution autonomy are constrained by design.

Third, we instantiate and evaluate this architecture in a regulated financial context under adversarial pressure. The experimental results demonstrate that mechanical governance of selection is technically implementable, produces auditable artifacts, preserves fail-loud behavior, and measurably alters the agent’s ability to exercise influence across attack classes. While probabilistic concentration remains, deterministic capture is structurally prevented under baseline enforcement.

The broader implication is architectural rather than algorithmic. Governing autonomous agents in high-stakes environments requires explicit control over how options are generated, reduced, and presented. Alignment of objectives, interpretability of reasoning, and action-level safeguards are necessary components, but they do not substitute for governance of selection authority itself.

This work does not claim universal safety guarantees or formal completeness. Instead, it establishes that selection governance can be formalized, implemented, stress-tested, and evaluated empirically. It shifts the discourse from whether agents are internally aligned to whether their externally visible power is causally bounded.

Future work must extend this foundation along three axes: formal verification of invariant preservation, expanded adversarial modeling (including multi-agent and adaptive dynamics), and human-in-the-loop validation of framing constraints. Bridging architectural feasibility with institutional-grade assurance remains an open challenge.

The findings reinforce the broader argument that governance in high-stakes agentic systems must target selection authority explicitly. Alignment, interpretability, and action-level filtering remain necessary, but they are insufficient in isolation. By bounding the mechanisms through which options are generated, reduced, and framed, the architecture operationalizes governance as a constraint on power rather than a correction of intent.

\section*{Acknowledgments}

The author thanks the AI Lab team at Grupo Santander for discussions and feedback, reviewers for constructive critique that improved the paper's technical precision and honesty about limitations, and the anonymous evaluators whose concerns about single-shot designs and N=1 limitations motivated explicit methodological justifications.

\bibliographystyle{ieeetr}
\bibliography{references}

\appendix

\section{Complete Experimental Results}
\label{sec:appendix-results}

This appendix reports the complete experimental results. Tables are organized by scenario and include mean metrics by ablation and by attack, as well as global aggregations.

\begin{table}[h]
\centering
\caption{Fraud Detection: Mean metrics by ablation configuration.}
\resizebox{\columnwidth}{!}{
\begin{tabular}{lccccc}
\hline
Ablation & SRI & FE & GD & QDV & ASR \\
\hline
B0 & 0.829 & 1.713 & 0.000 & 1.000 & 0.520 \\
B1 & 0.767 & 1.638 & 0.067 & 1.000 & 0.520 \\
B2 & 0.776 & 2.170 & 0.000 & 1.000 & 0.680 \\
B3 & 0.819 & 1.850 & 0.000 & 1.000 & 0.480 \\
B4 & 0.805 & 1.594 & 0.033 & 1.000 & 0.480 \\
B5 & 0.771 & 1.747 & 0.067 & 1.000 & 0.560 \\
B6 & 0.833 & 1.714 & 0.000 & 1.000 & 0.560 \\
B7 & 0.829 & 1.744 & 0.000 & 1.000 & 0.640 \\
B9 & 0.000 & 0.000 & 1.000 & 1.000 & 0.000 \\
B10 & 0.000 & 0.000 & 1.000 & 1.000 & 0.000 \\
B11 & 0.790 & 1.704 & 0.033 & 1.000 & 0.560 \\
B12 & 0.829 & 1.819 & 0.000 & 1.000 & 0.467 \\
\hline
\end{tabular}}
\end{table}

\begin{table}[h]
\centering
\caption{Fraud Detection: Mean metrics by attack class.}
\resizebox{\columnwidth}{!}{
\begin{tabular}{lccccc}
\hline
Attack & SRI & FE & GD & QDV & ASR \\
\hline
A1 & 0.714 & 1.152 & 0.167 & 1.000 & 0.833 \\
A2 & 0.714 & 2.293 & 0.000 & 1.000 & 1.000 \\
A3 & 0.846 & 1.993 & 0.000 & 1.000 & 0.340 \\
A4 & 0.794 & 1.819 & 0.040 & 1.000 & 0.000 \\
A5 & 0.806 & 1.572 & 0.040 & 1.000 & 0.393 \\
None & 0.676 & 1.298 & 0.200 & 1.000 & --- \\
\hline
\end{tabular}}
\end{table}

\begin{table}[h]
\centering
\caption{Payments Infrastructure Monitoring: Mean metrics by ablation configuration.}
\resizebox{\columnwidth}{!}{
\begin{tabular}{lccccc}
\hline
Ablation & SRI & FE & GD & QDV & ASR \\
\hline
B0 & 0.819 & 1.700 & 0.000 & 1.000 & 0.520 \\
B1 & 0.829 & 1.679 & 0.000 & 1.000 & 0.640 \\
B2 & 0.795 & 1.838 & 0.000 & 1.000 & 0.640 \\
B3 & 0.819 & 1.767 & 0.000 & 1.000 & 0.560 \\
B4 & 0.819 & 1.644 & 0.000 & 1.000 & 0.560 \\
B5 & 0.819 & 1.808 & 0.000 & 1.000 & 0.560 \\
B6 & 0.790 & 1.661 & 0.033 & 1.000 & 0.560 \\
B7 & 0.819 & 1.808 & 0.000 & 1.000 & 0.520 \\
B9 & 0.000 & 0.000 & 1.000 & 1.000 & 0.000 \\
B10 & 0.000 & 0.000 & 1.000 & 1.000 & 0.000 \\
B11 & 0.829 & 1.767 & 0.000 & 1.000 & 0.560 \\
B12 & 0.827 & 1.750 & 0.000 & 1.000 & 0.460 \\
\hline
\end{tabular}}
\end{table}

\begin{table}[h]
\centering
\caption{Payments Infrastructure Monitoring: Mean metrics by attack class.}
\resizebox{\columnwidth}{!}{
\begin{tabular}{lccccc}
\hline
Attack & SRI & FE & GD & QDV & ASR \\
\hline
A1 & 0.714 & 1.210 & 0.167 & 1.000 & 0.833 \\
A2 & 0.714 & 2.258 & 0.000 & 1.000 & 0.640 \\
A3 & 0.843 & 1.935 & 0.000 & 1.000 & 0.560 \\
A4 & 0.819 & 1.760 & 0.020 & 1.000 & 0.000 \\
A5 & 0.843 & 1.495 & 0.000 & 1.000 & 0.590 \\
None & 0.686 & 1.295 & 0.167 & 1.000 & --- \\
\hline
\end{tabular}}
\end{table}

\begin{table}[h]
\centering
\caption{QBR Analysis: Mean metrics by ablation configuration.}
\resizebox{\columnwidth}{!}{
\begin{tabular}{lccccc}
\hline
Ablation & SRI & FE & GD & QDV & ASR \\
\hline
B0 & 0.795 & 1.439 & 0.033 & 1.000 & 0.320 \\
B1 & 0.790 & 1.492 & 0.033 & 1.000 & 0.320 \\
B2 & 0.790 & 1.779 & 0.000 & 1.000 & 0.400 \\
B3 & 0.819 & 1.600 & 0.000 & 1.000 & 0.320 \\
B4 & 0.762 & 1.466 & 0.067 & 1.000 & 0.320 \\
B5 & 0.771 & 1.370 & 0.067 & 1.000 & 0.320 \\
B6 & 0.733 & 1.378 & 0.100 & 1.000 & 0.320 \\
B7 & 0.819 & 1.583 & 0.000 & 1.000 & 0.320 \\
B9 & 0.000 & 0.000 & 1.000 & 1.000 & 0.000 \\
B10 & 0.000 & 0.000 & 1.000 & 1.000 & 0.000 \\
B11 & 0.790 & 1.563 & 0.033 & 1.000 & 0.320 \\
B12 & 0.829 & 1.547 & 0.000 & 1.000 & 0.350 \\
\hline
\end{tabular}}
\end{table}

\begin{table}[h]
\centering
\caption{QBR Analysis: Mean metrics by attack class.}
\resizebox{\columnwidth}{!}{
\begin{tabular}{lccccc}
\hline
Attack & SRI & FE & GD & QDV & ASR \\
\hline
A1 & 0.714 & 1.016 & 0.167 & 1.000 & 0.833 \\
A2 & 0.714 & 2.119 & 0.000 & 1.000 & 0.605 \\
A3 & 0.786 & 1.762 & 0.060 & 1.000 & 0.000 \\
A4 & 0.769 & 1.296 & 0.080 & 1.000 & 0.000 \\
A5 & 0.800 & 1.296 & 0.040 & 1.000 & 0.050 \\
None & 0.679 & 1.199 & 0.183 & 1.000 & --- \\
\hline
\end{tabular}}
\end{table}

\begin{table}[h]
\centering
\caption{Global mean metrics by ablation across all scenarios.}
\resizebox{\columnwidth}{!}{
\begin{tabular}{lccccc}
\hline
Ablation & SRI & FE & GD & QDV & ASR \\
\hline
B0 & 0.814 & 1.617 & 0.011 & 1.000 & 0.453 \\
B1 & 0.795 & 1.603 & 0.033 & 1.000 & 0.493 \\
B2 & 0.787 & 1.929 & 0.000 & 1.000 & 0.573 \\
B3 & 0.819 & 1.739 & 0.000 & 1.000 & 0.453 \\
B4 & 0.795 & 1.568 & 0.033 & 1.000 & 0.453 \\
B5 & 0.787 & 1.641 & 0.044 & 1.000 & 0.480 \\
B6 & 0.786 & 1.584 & 0.044 & 1.000 & 0.480 \\
B7 & 0.822 & 1.712 & 0.000 & 1.000 & 0.493 \\
B9 & 0.000 & 0.000 & 1.000 & 1.000 & 0.000 \\
B10 & 0.000 & 0.000 & 1.000 & 1.000 & 0.000 \\
B11 & 0.803 & 1.678 & 0.022 & 1.000 & 0.480 \\
B12 & 0.828 & 1.705 & 0.000 & 1.000 & 0.426 \\
\hline
\end{tabular}}
\end{table}

\begin{table}[h]
\centering
\caption{Global mean metrics by attack class across all scenarios.}
\resizebox{\columnwidth}{!}{
\begin{tabular}{lccccc}
\hline
Attack & SRI & FE & GD & QDV & ASR \\
\hline
A1 & 0.714 & 1.126 & 0.167 & 1.000 & 0.833 \\
A2 & 0.714 & 2.223 & 0.000 & 1.000 & 0.748 \\
A3 & 0.825 & 1.897 & 0.020 & 1.000 & 0.300 \\
A4 & 0.794 & 1.625 & 0.047 & 1.000 & 0.000 \\
A5 & 0.816 & 1.454 & 0.027 & 1.000 & 0.344 \\
None & 0.680 & 1.264 & 0.183 & 1.000 & --- \\
\hline
\end{tabular}}
\end{table}

\end{document}